\documentclass[useAMS,usenatbib]{mn2e}
\usepackage{epsfig,graphicx,lscape}
\title{Automated Detection and Tracking of Solar Magnetic Bright Points}
\author[P. J. Crockett, D. B. Jess, M. Mathioudakis, F. P. Keenan]
       {P. J. Crockett, D. B. Jess, M. Mathioudakis, F. P. Keenan \\
        Astrophysics Research Centre, School of Mathematics and Physics, Queen's University, Belfast, BT7~1NN, 
Northern Ireland, U.K.}
\date{Accepted 2009 May 18.
      Received 2009 May 7;
      in original form 2009 March 3}

\pagerange{\pageref{firstpage}--\pageref{lastpage}}
\pubyear{2009}

\begin{document}

\maketitle

\label{firstpage}

\begin{abstract}
Magnetic Bright Points (MBPs) in the internetwork are among the smallest  objects in the solar photosphere and appear bright against the ambient environment. An algorithm is presented that can be used for the automated detection of the MBPs in the spatial and temporal domains. The algorithm works by mapping the lanes through intensity thresholding. A compass search, combined with a study of the intensity gradient across the detected objects, allows the  disentanglement of MBPs from bright pixels within the granules. Object growing is implemented to account for any pixels that might have been removed when mapping the lanes. The images are stabilized by locating long-lived objects that may have been missed due to variable light levels and seeing quality. Tests of the algorithm employing data taken with the Swedish Solar Telescope (SST),  reveal that $\approx90$\% of MBPs within a 75$^{\prime\prime}$ x 75$^{\prime\prime}$ field of view are detected. 
\end{abstract}

\begin{keywords}
{Sun: photosphere --- Sun: magnetic fields --- Sun: granulation --- techniques: image processing}
\end{keywords}

\section{Introduction}
\label{intro}
Magnetic Bright Points (MBPs) in the solar photosphere were first reported by Dunn \& Zirker \ (1973), who noted the presence of bright mottles that appear to break up into a network of grains''. These ``mottles'' were observed to move  distances comparable to granule diameters and some were clearly located in the inter-granular lanes (Dunn, Mann \& Simon \ 1973). MBPs are believed to be the foot points of magnetic flux tubes in the solar photosphere. Further investigations revealed that these magnetic concentrations correspond to kilogauss fields that are almost perpendicular to the solar surface (e.g. Stenflo \ 1985, Solanki \ 1993, S{\'a}nchez Almeida \& Martinez Pillet \ 1994).  

The magnetic network, a web-like structure of high magnetic field, forms at the edges of the super-granular cells and is believed to arise from longtime advection of magnetic flux to the perimeter of the supergranules (Rezaei et al. \ 2007). On the other hand, the internetwork is the area situated between the magnetic network boundaries. MBPs occur across the entire solar disk, appearing more numerous and dynamic near magnetic structures such as sunspots, whilst nevertheless existing within the less active internetwork regions. The identification and measurement of internetwork MBPs was pioneered by Muller \& Roudier (1984, 1992).  However, with the advancement of high resolution detection systems in recent years, the potential to understand these very small magnetic structures has significantly increased.  

It was originally thought that the internetwork was devoid of magnetic field.  However, Dom{\'{\i}}nguez Cerde{\~n}a, S{\'a}nchez Almeida \& Kneer \ (2003) found an average field of $\approx20$~G covering $\approx60$\% of the internetwork located mainly in the inter-granular lanes. When integrated over the solar disc, this provides a large fraction of the unsigned magnetic flux in the solar atmosphere.Through the use of spectropolarimetry it was revealed that the internetwork also includes areas of kilogauss fields which coincide with the positions of MBPs within this region (e.g. Grossmann-Doerth, Keller \& Schuessler \ 1996, S{\'a}nchez Almeida \& Lites \ 2000, Socas-Navarro \& S{\'a}nchez Almeida \ 2002). The large magnetic field strengths of the MBPs, compared to the surrounding field, make them a significant carrier for the internetwork flux (S{\'a}nchez Almeida et al. \ 2004). The association of MBPs with areas of increased magnetic field has been investigated by Ishikawa et al.  (2007) who conclude that efficient heat transport is also required to make the objects bright. More recently, de Wijn et al. (2007) have used Hinode magnetograms to study the dynamics of MBPs. An appreciation of how these small-scale kiloGauss objects are formed will  further our knowledge of how magnetic flux emerges from the interior to the surface of the Sun and subsequently evolves in space and time. It will also enhance our understanding of how larger magnetic objects form.  

In this paper we present a methodology for the automated detection and tracking  of the MBPs in the inter-granular lanes. \S~2 outlines the main properties of MBPs and the difficulties associated with their automated detection. A brief description of the observations and data reduction procedures is given in \S~3. \S~4 describes in detail how the algorithm works and the difficulties outlined in \S~2 are overcome. The detection rates of the algorithm are discussed in \S~5, while \S~6 summarizes our conclusions and presents an outlook for future work. 

\section{Properties And Challenges}
\label{Properties}
The two most notable properties of MBPs are their small size and increased brightness (Fig~1.).
\begin{figure}
\includegraphics[scale=0.8]{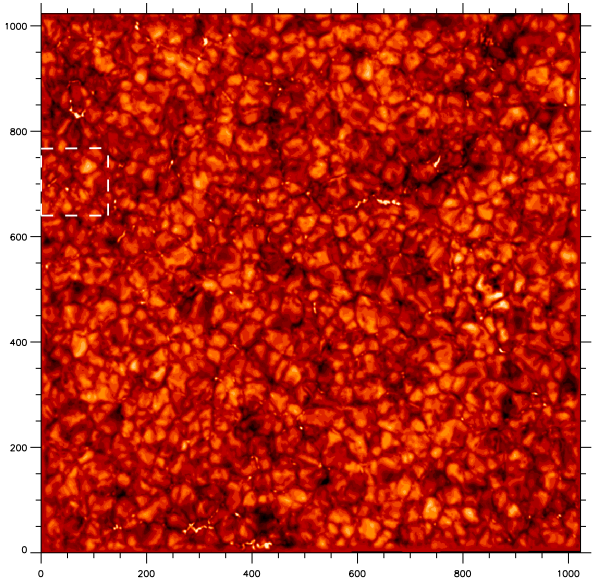} 
\caption{A 75$^{\prime\prime}$ x 75$^{\prime\prime}$ high spatial resolution, ($\approx50$km pixel$^{-1}$), image of solar granulation obtained with the Swedish Solar Telescope. A large number of MBPs can be identified in the image. These retain a high intensity compared to the photospheric background and are located within the inter-granular lanes.  They are the smallest objects resolvable by current optical telescopes, having an average diameter of $\approx200$km. The dashed white lines indicate a subsection to whcih the algorithm is apllied in \S~4. Tickmarks are pixels.}
\end{figure}
Individual MBPs rarely exceed diameters of $\approx300$~km with typical dimensions of $\approx150-250$~km (Berger et al.\ 1995, 2005). Single MBPs may merge together to form chains or groups several hundred kilometers long, but there appears to be no obvious correlation between size and brightness.  MBP intensities range from $\approx$ 0.8 - 1.8 times the mean photospheric value (S{\'a}nchez Almeida et al. \ 2004). The same authors investigated the surface area of the solar disk occupied by MBPs, calculating a coverage density of 0.3 MBPs per Mm$^{2}$.  The lifetime of individual MBPs are, for the most part, on a par with granulation, with the majority being $<10$mins, although some have been observed to survive for $>15$mins (M{\"o}stl et al.\ 2006) whilst groups of MBPs have been observed with lifetimes up to  $\approx75$~minutes (Berger et al \ 1998).  Velocities are projected to be in the region $0.5 - 5$~km~s$^{-1}$ and are believed to be driven primarily by the motions of the granulation (Berger \& Title \ 1996). They have magnetic field strengths in the range of 500G - 1400G (Beck et al.\ 2007).  All MBPs reside in the inter-granular lanes, where the  magnetic field concentrates due to the horizontal convective motions of the granulation and are never observed to exist within granules. Some MBPs are observed to form on the edge of granules, but are thought to be created by convective braking due to the down-flow within the lanes and hence are considered to be of non-magnetic origin (Berger \& Title \ 2001). MBPs appear as individual objects but through their evolutionary path they can change in a variety of ways (Berger et al. \ 2005). Shape deformation, such as elongation, due to the convective activity of the granule flow, is common for MBPs. Another frequent occurrence is the merging of small MBPs to form larger magnetic flux elements.  This merging process tends to occur at the intersection of several granules (Berger \& Title \ 1996) and creates chains or groups of MBPs.  The evolutionary properties of these groups varies somewhat to that of individual MBPs as they appear to have their motion restricted, with an average velocity a factor of 3 less than individual objects (Nisenson et al. \ 2003). The splitting of MBPs is thought to be the result of hydrodynamic shearing by photospheric flows and is reproduced in the numerical simulations of Carlsson et al. (2004).

The MBPs properties outlined above pose some unique challenges when attempting to identify them by means of an automated algorithm.  Their very small spatial scale is one of the main difficulties, as it is at the limit of our current spatial resolution.  Moreover, handling so few pixels increases the difficulty in stabilizing the dataset and identifying the same structure in successive frames, as variations in atmospheric seeing can cause the structures to often disappear. The large range of their intensities, 0.8 - 1.8 times the mean photospheric value, combined with intensity variations during their lifetime, makes the identification of MBPs with intensity techniques alone very difficult  and prone to errors.  Their relatively short lifetimes and rapid evolution, compared to larger magnetic structures, requires high cadence imaging over  extended periods of good seeing to allow entire life cycles to be observed. The limited density coverage dictates the need for a large field-of-view in order to pick up a significant number of these objects creating the essential need to remove large sections of data to identify MBPs only.  

A concerted effort has been made in recent years to develop tracking algorithms for solar magnetic structures. DeForest et al. (2007) compared four magnetic feature tracking codes by applying them to the same set of data and evaluate the circumstances under which each technique performs best. Several suggestions are made in the areas of data pre-processing, object identification, object association, object tabulation and event identification.  These recommendations will be considered further in section \S~4.6.

Previous detection algorithms for MBPs include a specialized version of the Multi-scale Pattern Recognition procedure by Bovelet \&~Wiehr \ (2007).  This procedure utilizes the intensity of MBPs to identify them via a four-stage process.  Firstly, the segmentation of all photospheric objects by setting equidistant intensity level thresholds from maximum to zero intensity is performed, producing a pattern of cells surrounding each local intensity maximum.  All the pixels within these cells are then intensity normalized to their cellular maximum. Next, the cells are shrunk to reasonable sizes by applying a single cut-off threshold to their normalized intensity profiles.  Finally, the cells are merged together removing the cellular pattern and recreating the individual objects.  This final step depends on the number of directly adjacent pixels between cells, whereby if there are insufficient adjacent pixels then the cells do not merge.   

Another relevant procedure known as the ``blob finding'' algorithm, was originally developed by Tomita \ (1990) and modified by Berger et al.\ (1995).  Within this, the following quantity is calculated for every pixel of the image:

\begin{equation}
B(x,y)=\frac{1}{(2M+1)^2}\sum_{u=x-M}^{x+m}\sum_{v=y-M}^{y+M} I(u,v)-\frac{1}{(2N+1)^2}\sum_{u=x-N}^{x+N}\sum_{v=y-N}^{y+N} I(u,v)
\end{equation}                                                                                                                      

where $M$, $N$ are integers, $M<N$ and $I(x,y)$ is the intensity of the pixel $(x,y)$, while the operator $B(x,y)$ returns either positive (bright blobs) or negative values (dark blobs). The resulting ``blob enhanced'' image is further processed with an unsharp mask algorithm to sharpen the boundaries of the bright structures.  A threshold operation is then performed, resulting in a binary image which has the value of unity at the locations of the bright points and zero elsewhere.  However some granulation peaks are also identified by this method and require elimination via dilation and erosion processes (Haralick, Sternberget \& Zhaung \ 1987) along with visual inspection.

\section{Observations}
\label{observations}
The data presented here are part of red continuum observations obtained on 2007 August 23, with the Swedish Solar Telescope (SST) on the island of La~Palma, using an optical setup as described in Jess et al.\ (2008).  Multi-Object Multi-Frame Blind Deconvolution (van Noort, van der Voort \& L{\"{o}}fdahl \ 2005) image restoration was implemented to remove small-scale atmospheric distortions from the data. Sets of 80 exposures were included in the restorations, producing an effective cadence of 9~s. All reconstructed images were subjected to a Fourier co-aligning routine where cross-correlation and squared mean absolute deviation techniques are utilized to provide sub-pixel co-alignment accuracy.  Image destretching, using a $40 \times 40$ correlation grid, was also implemented to remove image warping.

\section{The Detection Algorithm}
\label{analy}
Having undergone the initial processing and reconstruction, the data were then passed through a 6-stage process that is outlined below.  It is important to note that due to computational limitations it is not possible to pass an entire $1024 \times 1024$~pixel$^{2}$ image through the detection process.  Instead, the images are split into 64 $128 \times 128$~pixel$^{2}$ areas, with each area normalized to its average intensity.  These areas are then individually processed and are recompiled to form the complete detection image which is shown in Fig~8 overlaying the original image.  The following procedural details shall concentrate on a single $128 \times 128$~pixel$^{2}$ area (i.e. the highlighted section in Fig~1).

\subsection{Mapping out the Lanes}
\label{mapping}
The inter-granular lanes are regions of down-flowing plasma and appear as long dark regions in Fig~2(a). A common property of all MBPs is that they are located within the inter-granular lanes. To utilize this property  as an identification tool, the location of the lanes must be determined and mapped.  Their intensity level shows minimal variation within a given frame and therefore allows intensity thresholding to be used as a reliable identification method.  The upper  thresholding limit is determined as the mean intensity minus $0.8\times$ the sigma value of each $128 \times 128$~pixel$^{2}$ region. In addition, to be identified as a lane pixel it must have at least two conjoining pixels that also fall below the intensity threshold.  All these low intensity, conjoining pixels are then placed into a binary image, as shown in Fig~2(b), mapping out the lanes in white, with the granulation and MBPs in black. From Fig~2(a) and Fig~2(b) it is evident that the size of the lanes has been overestimated.  The overestimation of the lanes is required in order to single out MBPs completely from the granulation and allow their identification later on in the process.  However, this can create some problems. Large granules may have a lot of fine structure, leading to dark regions within granules. These regions can be mistakenly identified as inter-granular lanes causing large granules to be split apart and recognized as several small objects by the algorithm. Furthermore, the overall size and shape of the MBPs may be slightly affected as pixels that may belong to them are removed as lanes. The implementation of some additional routines can rectify these problems as detailed below. 

\subsection{Inverting the Lanes}
\label{inverting}
This is a relative short but necessary part of the process. The binary image from Fig~2(b) is inverted to produce an image of MBPs and granulation in white and the lanes in black, shown in Fig~3(a). A comparison of  Fig~2(a) and Fig~3(a) reveals that by overestimating the size of the lanes, we have separated the MBPs from the granulation, leaving them surrounded entirely by inter-granular lanes.

\begin{figure*}
\includegraphics[scale=0.05]{figures/fig2a.eps}\includegraphics[scale=0.05]{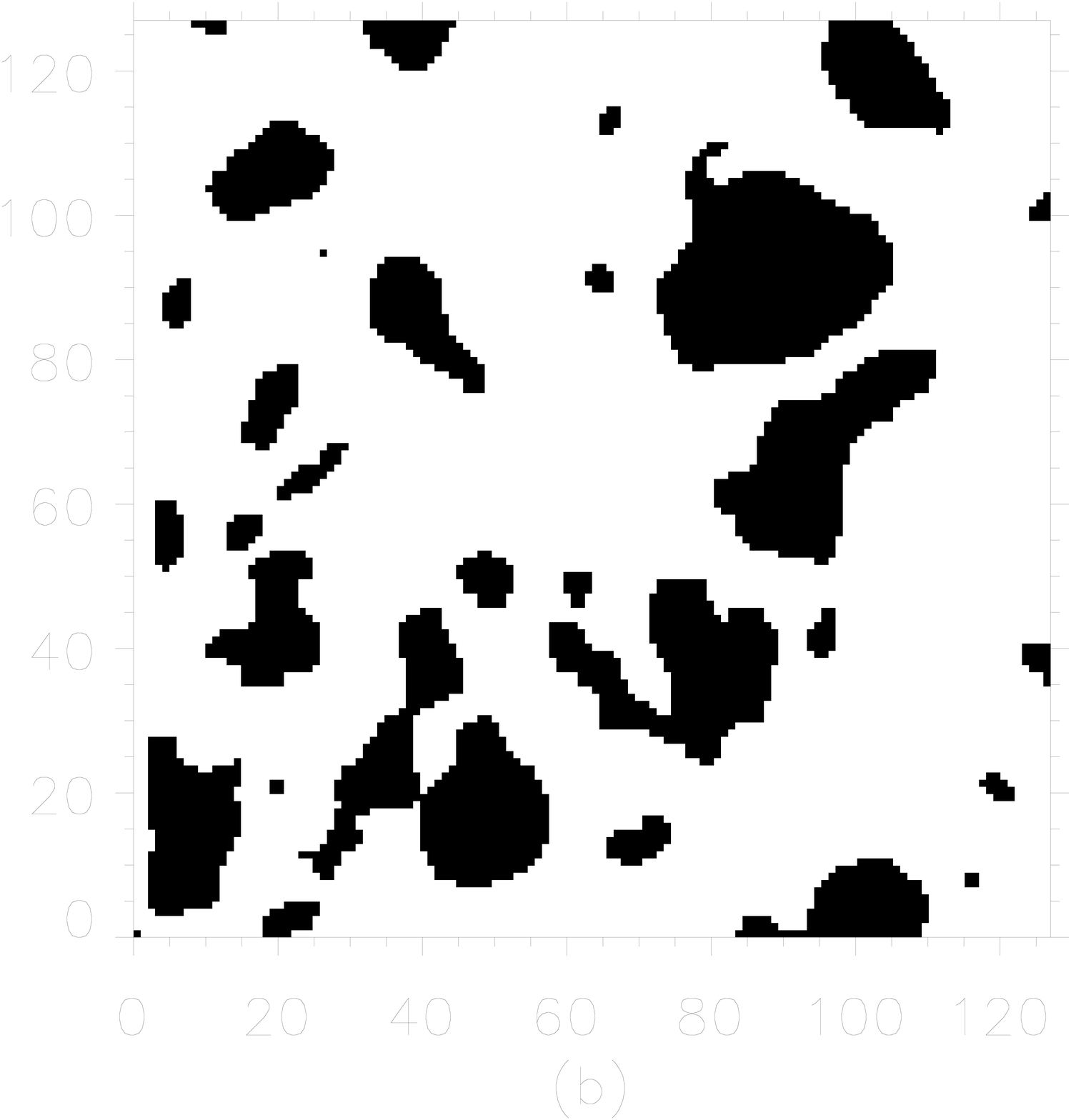}
\caption{(a) The highlighted $128 \times 128$~pixel$^{2}$ box in Fig~1. (b) A binary image of the lanes (white) along with the granules and MBPs (black)  The lanes are identified by setting an intensity threshold limit, and their size has been overestimated, resulting in a complete separation of the MBP from the granules. Tickmarks are pixels.}
\end{figure*}

\begin{figure*}
\includegraphics[scale=0.05]{figures/fig3a.eps}\includegraphics[scale=0.05]{figures/fig3b.eps}
\caption{(a) The inverted image. (b) The outcome of the compass search phase.  This examination stipulates that to be considered as an MBP pixel, it must be located within 7 pixels ($\approx350$~km) of a lane in all compass directions (N,S,E,W).  If a pixel fails to meet this criteria it is discarded. Comparison of Fig~3(a) and Fig~3(b) shows that the large objects have been removed.  Other small objects, such as exploding granules or large granules with some internal intensity structure, are also identified and produce false noise detections. Tickmarks are pixels.}
\end{figure*}

\subsection{Compass Search}
\label{compass}
This step in the procedure attempts to remove the large granulation structures.  The location of MBPs within the inter-granular lanes, combined with a data sampling of $\approx50$~km per pixel, allows us to place the following condition on the pixels of Fig~3(a).  Any pixel considered to be a MBP must be within 7 pixels ($\approx350$~km) of a lane in all directions of the compass, i.e. North, South, East and West.  Any pixels that do not comply with this compass search are discarded as pixels within a granule.  Due to their size, granulation pixels cannot be within 7 pixels of a lane in all directions and are thereby removed, whilst the MBP pixels are surrounded by lanes and are retained.  Fig~3(b) displays the result of this step in the procedure when it performed on Fig~3(a).  One of the main strengths of this approach is that it removes the non-MBPs that are situated at the edge of granules. However, utilizing the size of MBPs to identify them creates some limitations. The most significant arises from the merging of MBPs to form groups, which can become larger than 7 pixels across, causing them to be discarded in the process. Their large size will make their confinement within inter-granular lanes difficult, and they can merge with the granulation. In addition, the overestimate of the lanes causes the separation of larger granules into small objects. The compass search can falsely identify these small objects as MBPs. The use of a compass style search to isolate granules from intergranular lanes has also been used in the feature tracking algorithm of Strous (1995).
 
\subsection{Intensity Gradient}
\label{gradient}
In this step the algorithm calculates the intensity gradient across the objects found by the compass search. Fig~4 displays average line intensity profiles across all false detections and possible MBPs that have been visually identified in Fig~1. False detections are normally created as a result of the low intensity areas within large granules, causing them to be split into smaller objects whenever the lanes are identified. The line in Fig~4., representing the average gradient across MBPs, has a very steep intensity gradient in its boundaries compared to a more gradual increase for the false detections (line Fig~4). For the objects identified by the compass search, the intensity gradient is determined from the original image along a line of ten pixels positioned symmetrically and rotated about the individual object's centre of gravity.

\begin{figure}
\includegraphics[height=150pt,width=230pt]{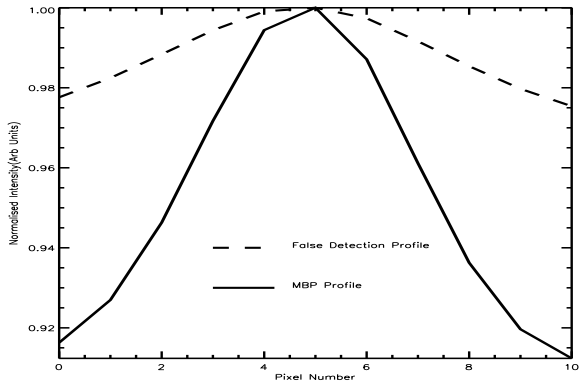}
\caption{Average light curves across MBPs and noise detected by the compass search.  Comparing the two lines verifies that MBPs generally obtain much steeper intensity gradients whilst false detections vary much more gradually.  Utilizing this outcome permits the removal of the false objects from the detection process.}
\end{figure}

The rotation of the line covers angles from $0^{\circ}$ (vertical) to $135^{\circ}$ in steps of $45^{\circ}$, and at each position the maximum rate of change in intensity, dI$_{max}$, is calculated. To be considered a MBP the object must retain a gradient greater than the set threshold in all four directions. The threshold is derived as the median intensity gradient of visually identified MBPs in the first frame of the time series minus 0.5 times their sigma value. The stipulation of varying direction within this process removes any false objects that may have been identified in close proximity to a MBP and would therefore have obtain a similar dI$_{max}$ in one given direction.  Fig~5. illustrates how the dI$_{max}$ of all the objects identified by the compass search from Fig~1. vary with angle. The red line on the graphs indicates the lower cut-off limit for MBPs, whilst the orange line marks the median dI$_{max}$ of MBPs. Fig~6. displays the resultant image from this process.     

\subsection{Growing}
\label{Growing}
Overestimation of the inter-granular lanes in the first step of this process  can lead to the reduction in the size and shape of the detected objects. In the final step of the algorithm the identified MBPs are grown to recover pixels that may have been removed in the first stage of this process. Growing operates by using the location of the identified bright points as ``seed'' regions.  An intensity range is then determined from the upper and lower intensities found at the positions of these seed regions in the original data. Any pixels immediately adjoint to the seed regions that possess an intensity within this range are then included in the final detection.  Fig~7(a) \&  Fig~7(b) show a comparison of the originally detected MBP and the finalized grown MBP, respectively. The contours in Fig~7(c) outline the perimeter of the finalized grown MBPs and demonstrates that they have been detected and grown to a high degree of accuracy. Finally, after passing through the five processes described above, each of the 64 $128 \times 128$~pixel$^{2}$ boxes are reassembled to form a binary detection frame of the entire field-of-view, where regions with MBPs are indicated (Fig~8).

\begin{figure*}
\includegraphics[scale=0.015]{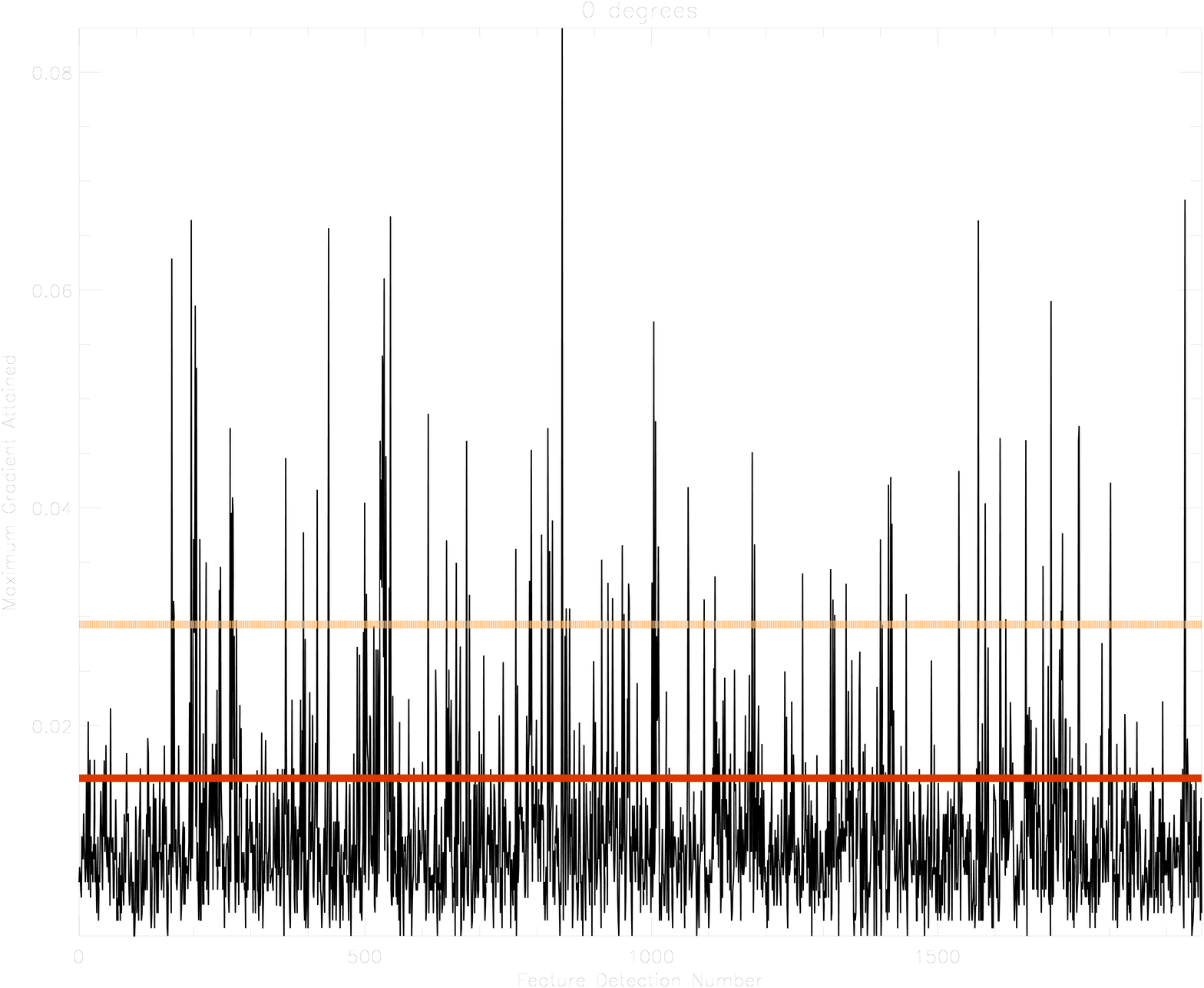}\includegraphics[scale=0.015]{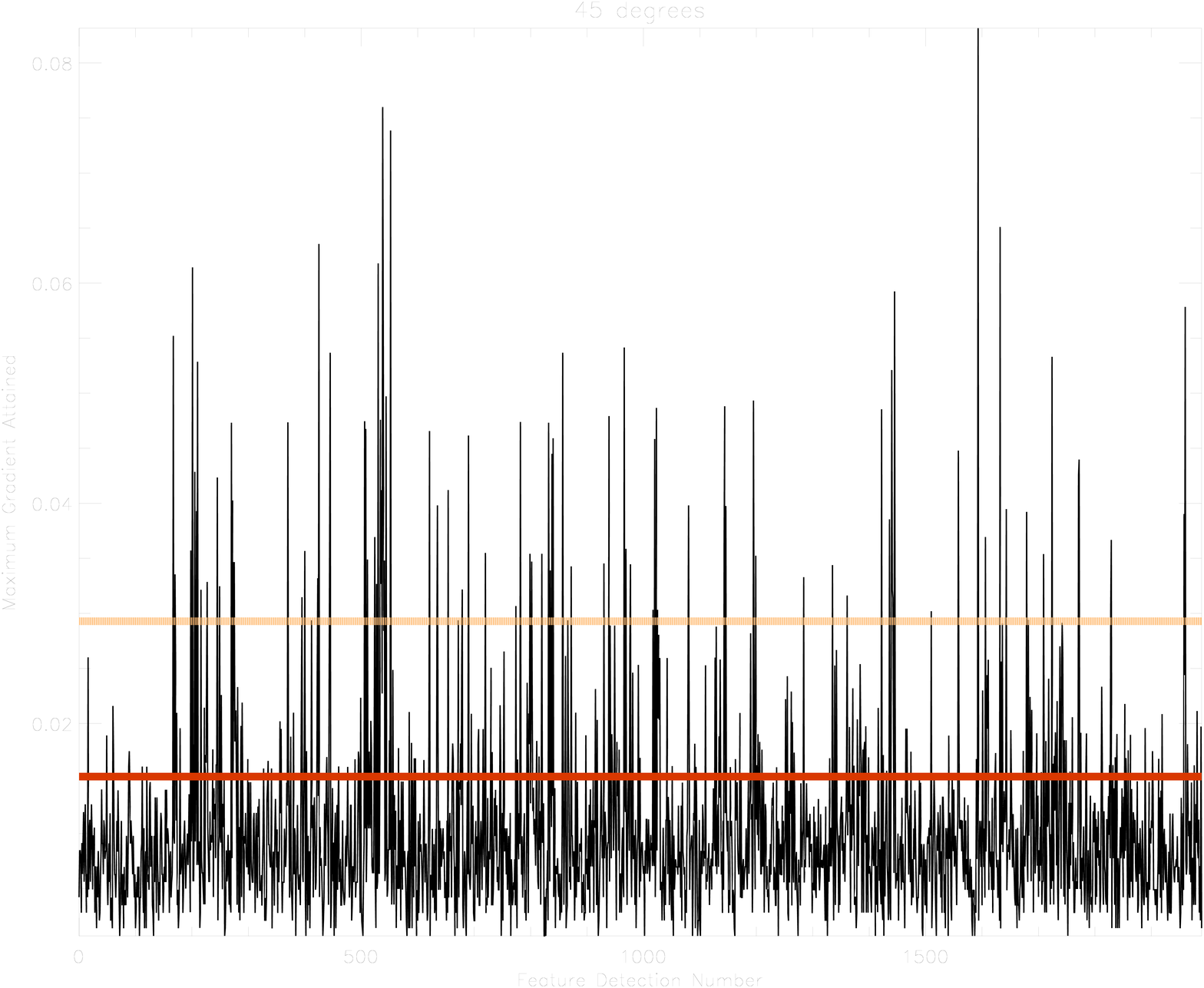}
\includegraphics[scale=0.015]{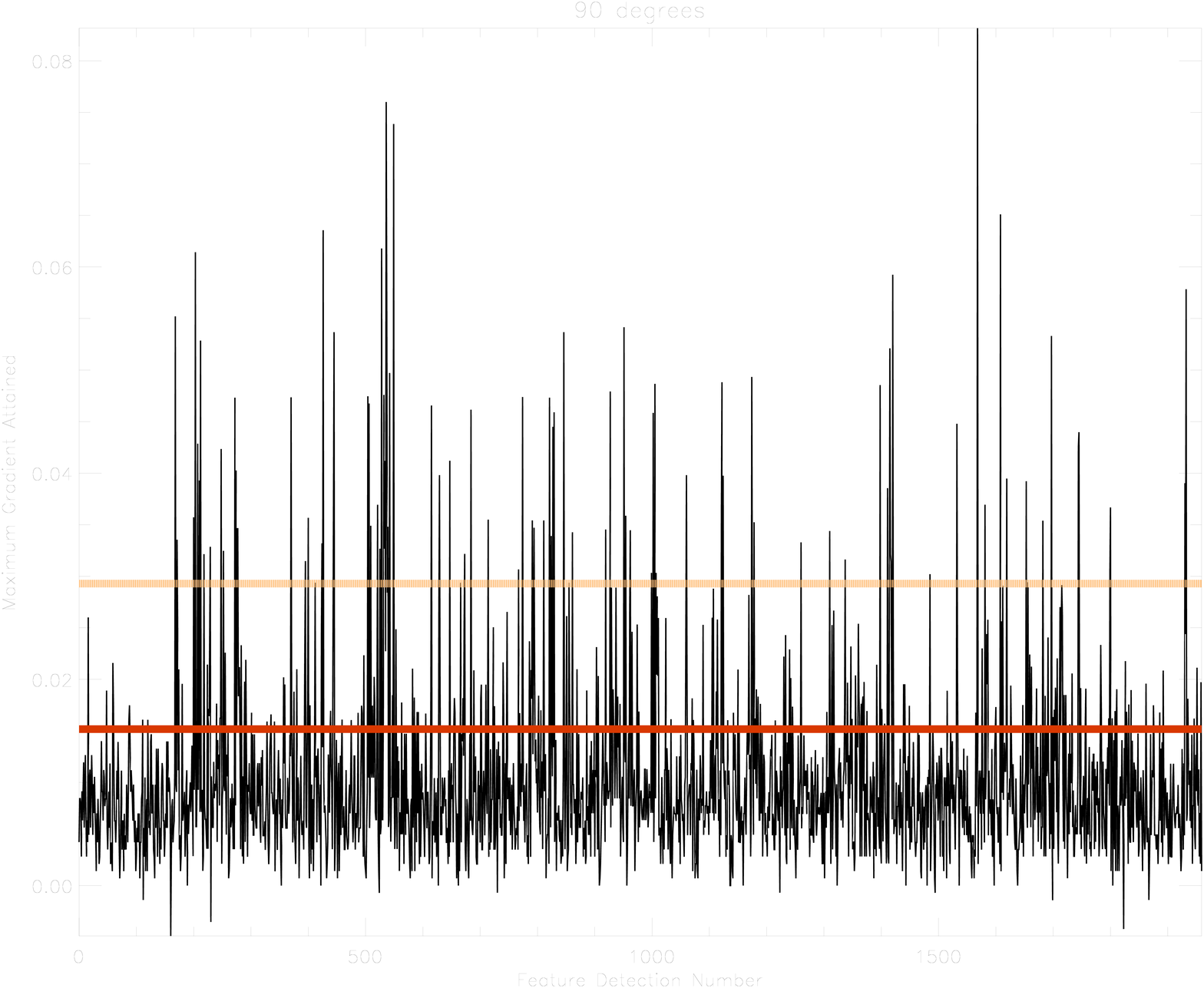}\includegraphics[scale=0.015]{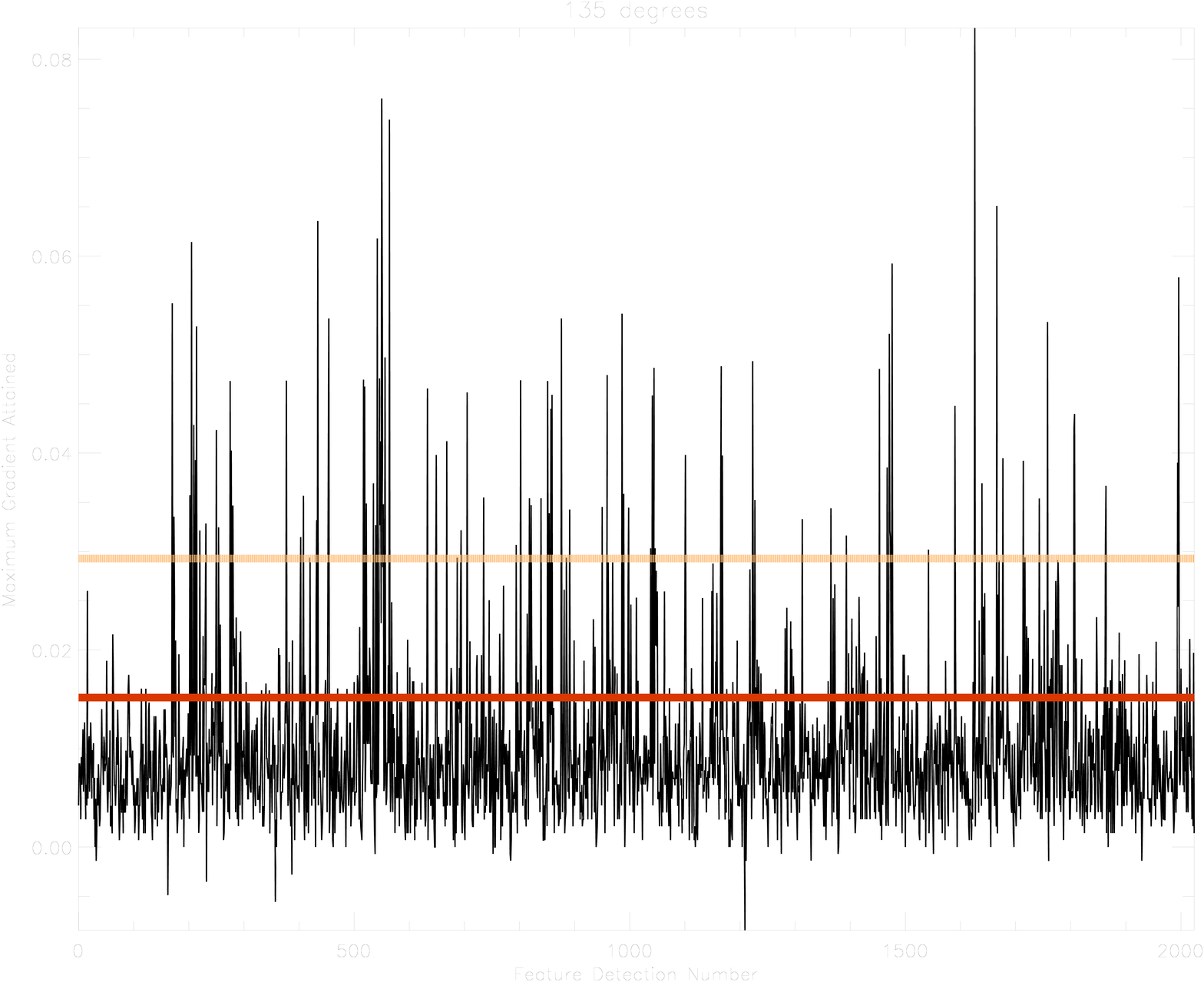}
\caption{The four plots show the maximum intensity gradients achieved by all the objects identified by the compass search over the four angles $0^{\circ}$,$45^{\circ}$,$90^{\circ}$,$135^{\circ}$.  The red line in the graphs represents the lower cut off gradient; objects that obtain an intensity gradient above this in all four graphs are identified as MBPs.  The orange line marks the median intensity gradient of MBPs.}
\end{figure*}

\begin{figure*}
\includegraphics{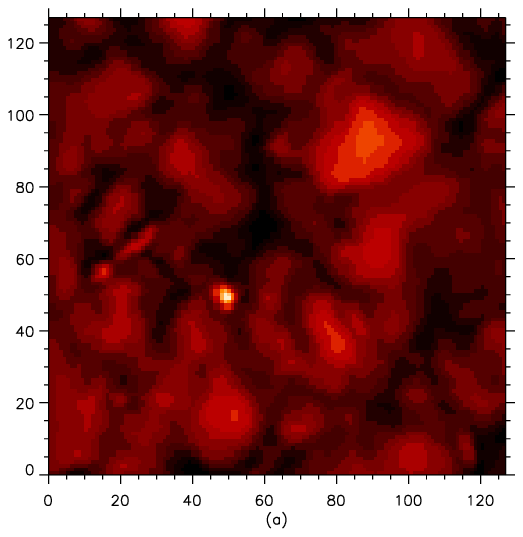}\includegraphics{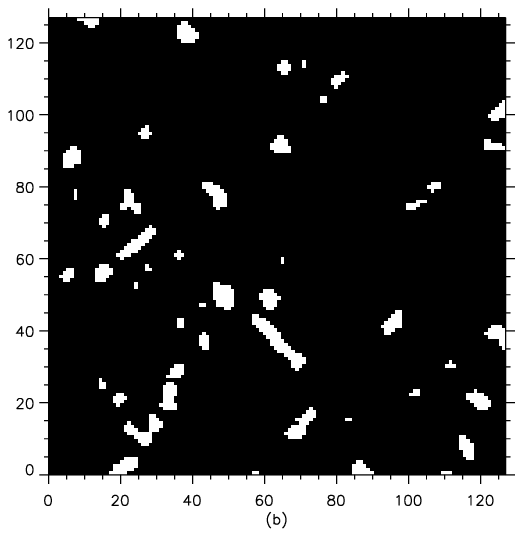}\includegraphics{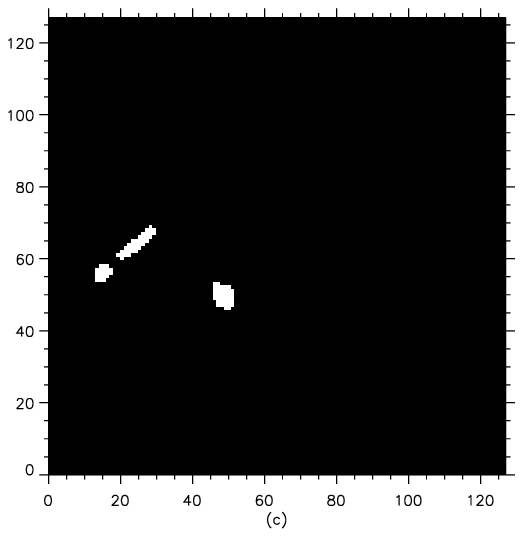}
\caption{(a) Original data (b) The outcome of the compass search (c) The outcome of the intensity gradient phase. Tickmarks are pixels.}
\end{figure*}

\begin{figure*}
\includegraphics{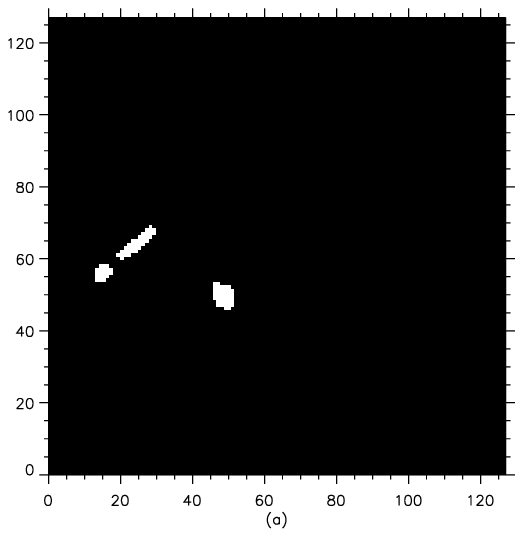}\includegraphics{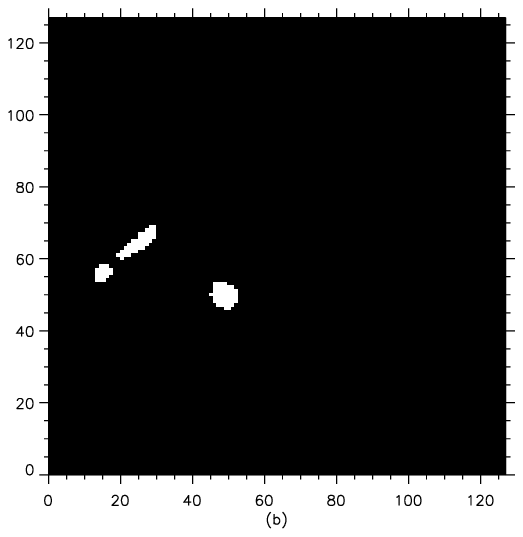}\includegraphics{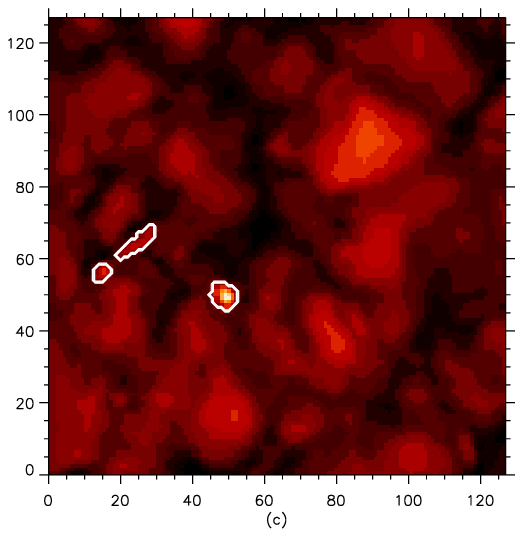}
\caption{(a) The detected MBPs. These may be slightly altered in size and shape due to the overestimation of the lanes in the first stage of the algorithm.  (b) Using the maximum and minimum intensity levels found in the seed regions in panel (a) the algorithm grows the MBPs to their correct size and shape. (c) The original image is over plotted with the contours outlining the perimeter of the detected MBPs.  From this it is possible to see the accuracy to which the MBPs are found. Tickmarks are pixels.}
\end{figure*}

\subsection{Stabilization And Tracking}
\label{stabilization}
Some MBPs may not be consistently recognized throughout their lifetime in successive binary detection frames, although they may exist in the original data.  This complication may be expected since a small variation in seeing quality can lead to significant variations in the intensity and contrast levels of the entire image, the most notable difference occurs in regions of high intensity (Title \& Berger \ 1996).  This stage in the process attempts to stabilize the results by establishing long-lived objects and locating these in frames where they may have been missed.  Moreover, it removes objects that are short-lived (i.e. a lifetime $<45$~s) from the detection process. The latter tend to be noise or poorly established MBPs.  Stabilization and tracking occurs in the stages described below and is illustrated in Fig~9. \& Fig~10.

We initially ascertain the first frame in which the MBP is detected,(frame {\it i}), followed by the calculation of its centre of gravity in this frame. We search around this centre of gravity in the succeeding, (frame {\it i+1}), for detected MBPs. The area over which this search is performed can be varied for different cadences or spatial sampling parameters. In the present case the search is performed over a $6 \times 6$~pixel$^{2}$ area symmetrically positioned about the centre of gravity, i.e. 3 pixels in all directions. Given typical MBP velocities (Berger \& Title \ 1996) and the 9s cadence of our data, the MBP movement will be limited to less than 2 pixels in any direction between successive frames. We therefore believe that this $6 \times 6$~pixel$^{2}$ area is justified.

Object association is the method by which an object in a succeeding frame, {\it (i+1)}, is associated with, or identified as, the same object from the previous frame {\it (i)}. If an associated MBP is found to exist in the succeeding frame, ({\it i+1}), the system does not require stabilization and the algorithm continues to the next frame. If no associated MBP is found in the succeeding frame, ({\it i+1}), the algorithm will attempt to stabilize the detection of that object.  Stabilization examines the following five frames in sequence ({\it (i+2):(i+6)}). The search is stopped if an associated MBP is discovered in any of these 5  frames.  A seed region can then be defined as the area where the MBP persists for the majority of the previous five frames ({\it (i-4):(i)}).  The MBP can then be grown in the frame where it was missing, using a procedure similar to the one described in \S~4.5. If the growth is greater than the average number of pixels that form the object within the previous five frames, then this value is set as a limit to the growth of the object.  The growth is fixed at the number of pixels closest to the centre of the seed region. Since the MBPs can have intensities below that of the  mean photosphere, the growth must be limited to prevent the inclusion of granules in these circumstances.  The grown MBP is then included in the detection results, thereby stabilizing the detection of the object.  If a confirmed MBP is not found within the five subsequent frames of the time series, it is concluded that it no longer exists.  Any short lived objects, those found not to exist in at least five consecutive frames, (i.e. lifetime $<45$~s), are discarded during the stabilization process as they tend to be noise. 

Tracking of the objects operates simultaneously with stabilization. A framework for the development of tracking systems, put forward by the DeForest et al. (2007), recommends the following best practices that may be followed for feature tracking applications; data pre-processing, object identification, object association, object tabulation and event identification. The first two stages have already been described in \S~3 and \S~4.1-\S~4.5. respectively.  The algorithm presented here has a three stage object association procedure. Firstly, it identifies and separates objects in the succeeding frame that are found to exist within the search area, created around the center of gravity of the object from the previous frame.  Secondly, it calculates the center of gravity for each of these objects.  If an object's center of gravity is found to exist outside the search area, it is immediately considered as a separate object.  On the contrary, if an object's centre of gravity exists within the search area, it is considered as a possible associated MBP.  Finally, if two or more objects are detected within the search area, the object with the closest center of gravity to the original shall be defined as the associated MBP. This method allows the centre of gravity of all individual MBPs to be tracked throughout their lifetime (see online material). Object tabulation is concerned with recording information about individual objects such as lifetimes, velocities, directional bias, location and area covered. This step is beyond the scope of this paper and will be studied in a subsequent publication.  

Event identification concerns the tracking of objects during the creation and demise phases of their lifetimes.  Creation of MBPs can occur as isolated appearance or fragmentation.  For isolated appearance, the detection algorithm will identify newly  emergening objects which can then be individually tracked.  Fragmentation, or splitting, of MBPs is a common occurrence with an average time between events of a few 100 seconds (Berger \& Title, \ 1996).  The tracking algorithm treats every MBP as a separate entity, tracking only one MBP at any one time. Consider a scenario where a single MBP, object {\it A}, exists in frame {\it (i)}.  In the following frame, {\it (i+1)}, the MBP has split into two separate objects namely {\it B} and {\it C}.  The detection algorithm will identify two separate objects, {\it B} and {\it C}, that are in close proximity to each other in frame {\it (i+1)}.  It would have also detected the single MBP, object {\it A}, in frame {\it (i)}. 

The tracking algorithm then investigates the search area in the frame {\it (i+1)}, surrounding the centre of gravity of object {\it A}.  There will be two separate objects present in this search area, object {\it B} and object {\it C}.  The algorithm will now recognize that the two objects exist in the search area and will treat them as separate.  The algorithm will continue by calculating the individual objects centre of gravity and will determine if both centers of gravity exist within the search area.  If only one center of gravity exists within the search area, e.g. {\it B} but not {\it C}, then object {\it B} is considered to be the continuation of MBP {\it A}. Object {\it C} is considered as a separate structure and is tracked as such.

 If the centers of gravity of both {\it B} and {\it C} exist within the search area, then the object that has the closest centre of gravity to that of object {\it A}, the original MBP in frame (i), is identified as a continuation of that feature. The other feature is treated as a separate MBP and shall be tracked as such.  If neither {\it B} or {\it C} exists within the search area, the original MBP, {\it A} is considered to no longer exist.  The same procedure can be applied to MBPs that split into more than two features. In summary, the component of a fragmented MBP that is closest to the center of gravity of the original MBP, is considered to be a continuation of the original object, whilst the other fragments are considered separately by the tracking algorithm.

The demise of MBPs can occur as isolated disappearance or merging. Once again the detection algorithm shall track the isolated case as it identifies whenever an object has disappeared, and hence, the tracking algorithm is not invoked.  Merging is more complicated as the MBPs do not disappear or cease to exist, but form a secondary structure.  Consider the scenario where  MBPs {\it A} and  {\it B}, exist in close proximity in frame {\it (i)}, while in frame {\it (i+1)}, only one MBP exists {\it C} (i.e. two objects have merged to form one). The tracking algorithm handles each detected MBP as a separate object. One of the two separate objects, {\it A} or {\it B}, shall be identified and tracked to the next frame {\it (i+1)}.  For example, object {\it A} is examined first by the tracking algorithm. The search area shall be placed around the centre of gravity of object {\it A} in the frame {\it (i+1)} and will find the single MBP, {\it C}, present in that frame. Object {\it C} will be accepted or rejected as the continuation of object  {\it A} accordingly. If accepted as the associated MBP, object {\it C} is removed from future searches of the area.  When the tracking algorithm examines object {\it B}, it will find no associated MBP in the following frame. Object {\it B} is therefore considered to no longer exist.  

\begin{figure}
\includegraphics[scale=0.8]{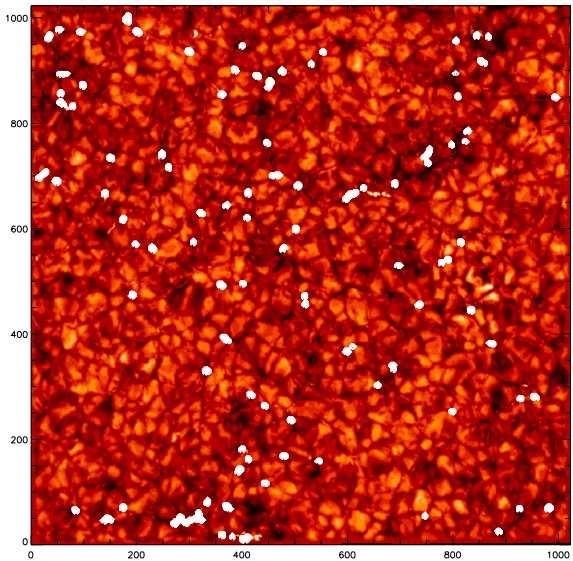}
\caption{After passing through the 5 step process described in\S~4 the 64 $128 \times 128$~pixel$^{2}$ boxes are reassembled to create a $1024 \times 1024$~pixel$^{2}$ binary detection frame.  The figure displays this binary image projected onto Fig~1.  Regions marked in white represent detected MBP positions. Tickmarks are pixels.}
\end{figure}

\section{Results and Discussion}
\label{results}
To determine the accuracy of the algorithm's detection rate, 10 successive data frames were examined and MBPs were identified in each $1024 \times 1024$  pixel$^2$ frame.  MBPs were primarily identified visually using stringent parameters, whereby there was a high degree of confidence that a object was a MBP. The algorithm was then applied to the data and the number of positive detections, number of false detections and the number of MBPs that were not detected was recorded.  A total of 1300 MBPs over all 10 frames were visually identified in this test. There were 1118 positive algorithm detections, with an average detection rate per frame of ~$\approx87$\%, leaving ~$\approx13$\% of the MBPs undetected. 
However, the overall number of detections across the 10 frames generated by the algorithm was 1474, producing 356 false detections, leading to an average false detection rate of ~$\approx23$\%.  This percentage may appear large, but on average these false detections occupy less than ~$\approx0.05$\% of the resultant binary image.  

The criteria employed to visually identify the MBPs were subsequently relaxed to take into account the full range of their intensity levels ranging from $0.8 - 1.8$ times the mean photosphere (S{\'a}nchez Almeida et al. \ 2004).  All objects that reached or surpassed this lower intensity limit of 0.8 times the mean photospheric value  were considered as MBPs, unless they were an obvious invalid detection. This approach produced a total of 1541 MBPs, and 1365 were identified as positive detections by the algorithm, a slight increase in the average detection rate per frame to ~$\approx89$\% and ~$\approx11$\% remaining undetected. However, there was a sharp decrease in false detections to an average rate of ~$\approx7$\%, amounting to 109 false detections over all 10 frames, with them now accounting  for ~$\approx0.01$\% of the binary image. 

The false detections are created for a number of reasons, the primary being large granules that have internal intensity structure and possess bright regions near the boundary with the inter-granular lanes.  Our algorithm can interpret the low intensity region of the granule as a lane causing the granule to split during the lane mapping process.  The bright edge is then identified by the compass search as a possible MBP structure as it is surrounded entirely by lanes.  Finally, and due to the close proximity of the bright edge to the inter-granular lane, the maximum gradient of the object is sufficient to pass as a MBP.  The probability of this sequence occurring has been minimised by stipulating that the intensity gradient must surpass the threshold in all 4 analyzed directions, thereby reducing the likelihood that long thin structures, comparable to bright granule edges, shall be accepted.  Exploding granules, and especially the latter stages of their evolution where they have been completely separated from their host granule, provide another source of incorrect detections. They produce small blobs of plasma that appear in the inter-granular lanes adding to the problem of identifying MBPs.  Usually these blobs tend to be of low intensity and the majority are abandoned by the algorithm at the intensity gradient stage of the process.

In some circumstances the procedure can miss brights point. For instance, very large MBPs and MBP groups may be removed by the algorithm during the compass 
\begin{figure*}
\includegraphics{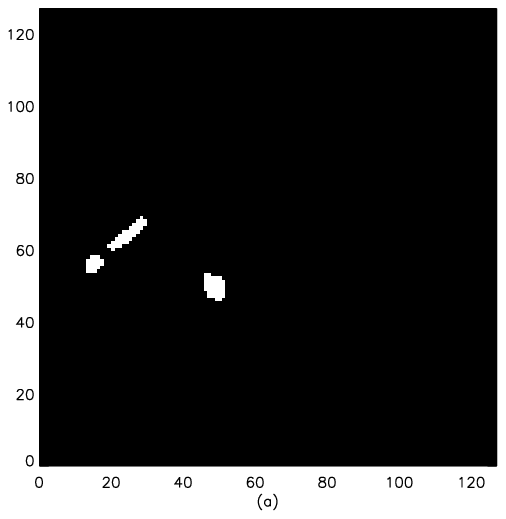}\includegraphics{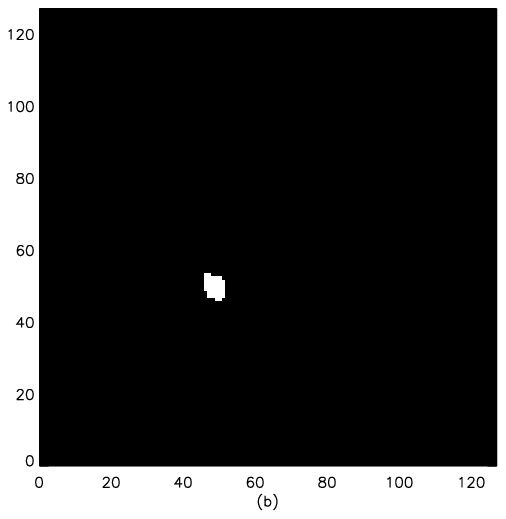}\includegraphics{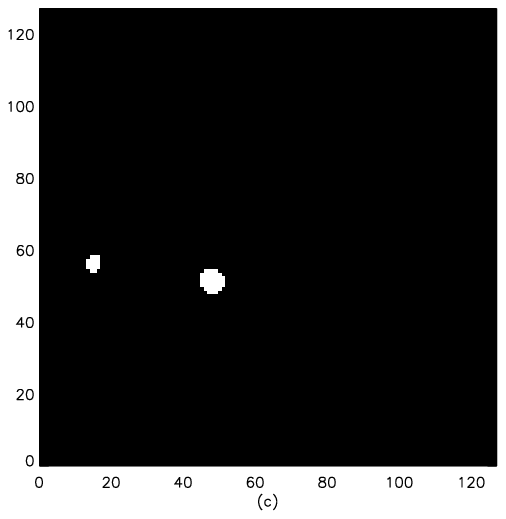}
\includegraphics{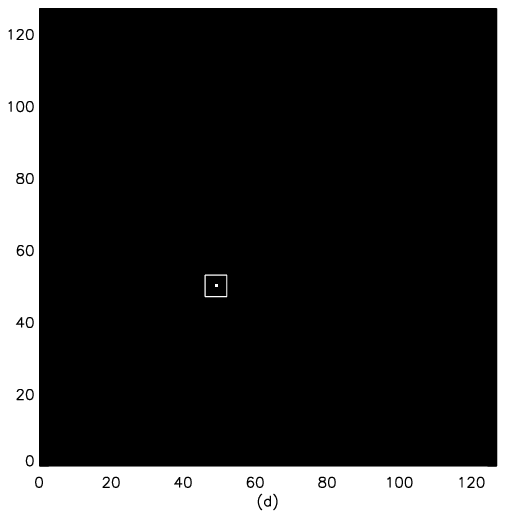}\includegraphics{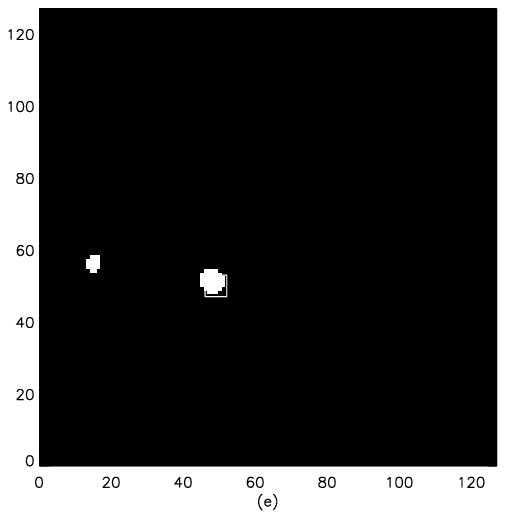}\includegraphics{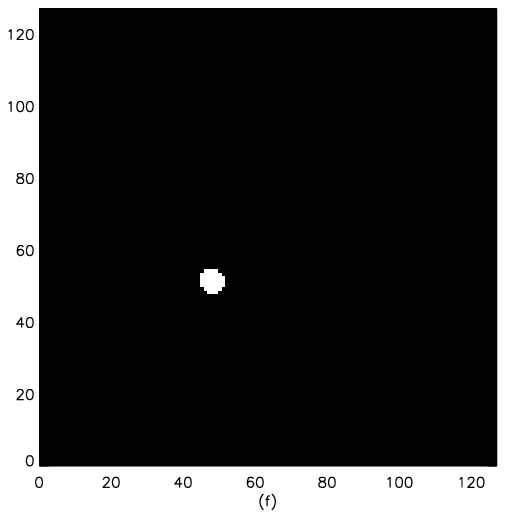}
\caption{A description of the tracking procedure when an associated MBP is located in the succeeding frame. (a) First frame where the MBP exists, {\it frame (i)}. (b) The single MBP under investigation from {\it frame (i)}. (c) The succeeding frame, {\it frame (i+1)}. A visual comparison of (b) and (c) reveals that an associated MBP is clearly detected by the algorithm in {\it frame (i+1)}. (d) The search area located around the centre of gravity of the MBP in {\it frame (i)}. (e) The MBP in the succeeding {\it frame (i+1)}, is shown to exist within the search area.  Importantly  the MBP's centre of gravity exists withing the search area and object association is confirmed.  (f) The MBP from the succeeding {\it frame (i+1)}, is then grown, included in the final result and tracking is hereby achieved.  Tickmarks are pixels.}
\end{figure*}
\begin{figure*}
\includegraphics{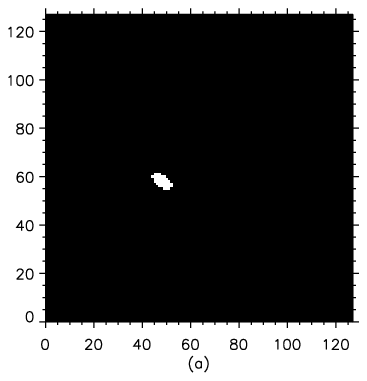}\includegraphics{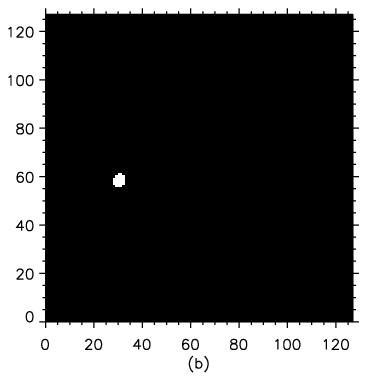}\includegraphics{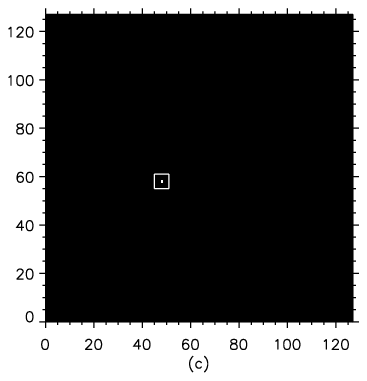}\includegraphics{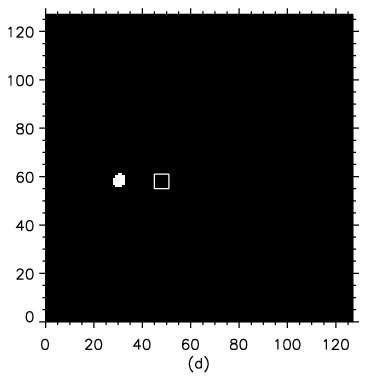}
\includegraphics{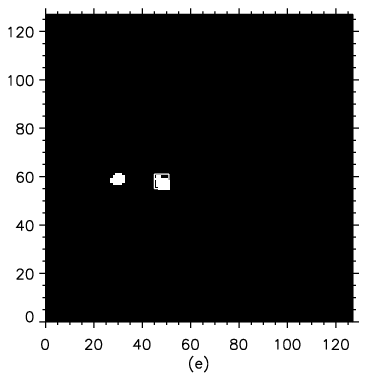}\includegraphics{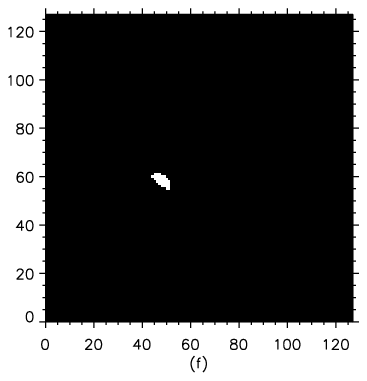}\includegraphics{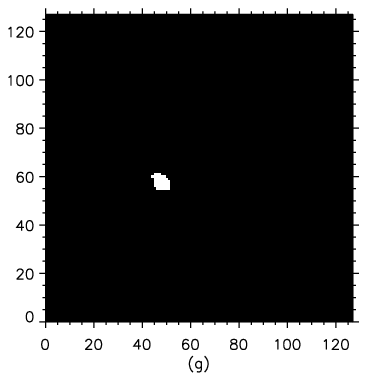}\includegraphics{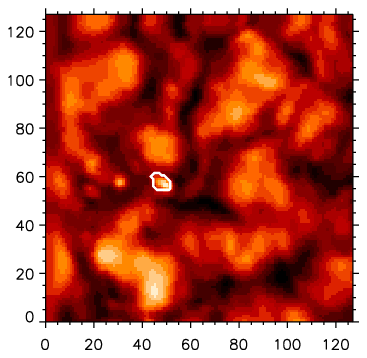}
\caption{Application of the stabilization procedure when a MBP is missing from the succeeding detection frame. (a) The preceding frame, {\it frame (i)}, where the detected MBP exists. (b) The succeeding frame, {\it frame (i+1)}.  The MBP which was detected to exist on the right of the previous image, {\it frame (i)},  is missing here. (c) The search area set around the centre of gravity of the MBP in {\it frame(i)}. (d) No MBPs exist within the search area in the succeeding {\it frame (i+1)}.  The algorithm therefore projects the search area forward to the following 5 frames, {\it frame(i+2):(i+6)}. (e) An associated MBP has been found within the search area in the following 5 frames, {\it frame(i+2):(i+6)}.  The stabilization procedure therefore shall now attempt to grow the MBP in the frame where it was not detected. If a MBP is not found to survive within the search area in the following 5 frames, then that object is considered to no longer exist. (f) The MBP has been located in the majority of the previous 5 frames, {\it frame(i-4):(i)} thereby defining a seed position. (g) The missed MBP, in {\it frame (i+1)},  is grown from this seed location using intensity thresholds generated from the maximum and minimum levels from the same area of the original image, {\it image (i+1)}. The grown MBP which is included in the final results and is thereby tracked by the algorithm. (h)  The {\it image (i+1)} from which the MBP was not detected.  The white contours mark the outline of the grown MBP.}
\end{figure*}
search as their scales are similar to small granules.  Elongated MBPs may be missed if they are aligned along one of the directions over which the intensity gradient is being calculated, as the intensity gradient does not change quickly over a MBP.  Finally, MBPs may be missed due to the poor definition of lane to granule boundaries.  Lanes are rarely well defined objects, with a slight haze of plasma overshooting from the granule on most occasions.  In some cases this haze  causes problems for the intensity gradient section of the algorithm.  If a MBP exist within a well defined lane then the rate of change in intensity, going from a dark region across the bright point, is markable and a very useful tool for identification.  However if the MBP is surrounded in the lane by overshooting plasma from the granules, then the intensity gradient may not be so steep, leading to a failure to attain the required threshold level.  Similar to this effect is the loss of MBPs that are in very close proximity to a neighboring granule, as once again the lane between granule and MBP may not be defined to a sufficiently high degree to permit the required rate of change in intensity to be discovered.     

\section{Concluding Remarks}
\label{conc}
The emergence and evolution of small kG magnetic fields on the solar surface is one of the most interesting topics in solar physics. The very small size of MBPs, at the limit of our spatial resolution, make them difficult to detect. Their large intensity range creates further difficulties, especially in terms of what exactly constitutes a MBP.  Their continuous evolution and short lifetimes combined with deteriorating observing conditions, can make meaningful investigations of their evolutionary properties difficult. Their relatively small density coverage causes data handling difficulties, as the majority of a large field of view must be discarded. The existence of non-magnetic bright points in close proximity to inter-granular lanes can produce false detections.\\

Our paper presents an algorithm that can be used for the automated detection and tracking of MBPs in the internetwork. An evaluation of the algorithm using observations from the Swedish Solar Telescope, reveals that $\approx90$\% of MBPs are identified with a false detection rate of $\approx10$\%. The false detections occur primarily during the mapping of the inter-granular lanes as the splitting of the granules can generate high contrast areas within a granule which are mistaken as an inter-granular lane. The introduction of a compass search creates a limitation to the size of MBPs that can be picked up by the algorithm using a 7 pixel radius, this correspondsto an artificial area limit of 370,000 km$^2$. MBPs which are larger than that are not considered. A dramatic change in the shape of an MBP can shift its center of gravity significantly, placing it outside the search area and rendering the algorithm unable to track it. Some of these difficulties may be overcome with a new generation of high cadence instruments such as Rapid Oscillations in the Solar Atmosphere (ROSA) and the Crisp Imaging SpectroPolarimeter (CRISP). Finally, we emphasize that although the algorithm has been developed and tested on red continuum imaging data, it can be directly applied to any high contrast datasets including dopplergrams and magnetograms.

\section*{Acknowledgments}
This work has been supported by the Science and Technology Facilities Council. 
The SST is operated on the island of La Palma by the Institute for Solar Physics of the Royal Swedish Academy of Sciences in the Spanish Observatorio del Roque de los  Muchachos of the Instituto de Astrofisica de Canarias. These observations have been funded by the Optical Infrared Coordination network, an international collaboration supported by the European Commission's Sixth Framework Programme. 
PJC would like to thank the European Social Fund for a PhD studentship. 
FPK is grateful to AWE Aldermaston for the award of a William Penney Fellowship. We would like tothank an anonymous referee for useful comments and suggestions.

\bsp

\label{lastpage}

\end{document}